\PassOptionsToPackage{table}{xcolor} % Daniel: to make it work with the ACM template which already specifies xcolor but without that option

\documentclass[acmsmall, screen=true, anonymous=false, nonacm]{acmart}

\usepackage{tabularx}
\usepackage{dcolumn} %Aligning numbers by decimal points in table columns
\newcolumntype{d}[1]{D{.}{.}{#1}}

\definecolor{rowcol}{rgb}{0.9,0.9,0.9}

\usepackage{todonotes}
\let\xtodo\todo
\renewcommand{\todo}[1]{\xtodo[inline,color=green!50]{#1}}

\newcommand{\change}[1]{#1}
\usepackage{subfigure}
\usepackage{multirow}
\AtBeginDocument{%
  \providecommand\BibTeX{{%
    \normalfont B\kern-0.5em{\scshape i\kern-0.25em b}\kern-0.8em\TeX}}}

\setcopyright{acmcopyright}
\copyrightyear{2024}
\acmYear{2024}
\acmDOI{XXXXXXX.XXXXXXX}

\acmConference[MobileHCI'24]{Make sure to enter the correct
  conference title from your rights confirmation emai}{September 30--October 3,
  2024}{Melbourne, Australia}
\acmPrice{15.00}
\acmISBN{978-1-4503-XXXX-X/18/06}

\begin{document}


%\title{Putting Mobile Text Inputs into Context: Three Research Patterns}
%\title{Context-Aware Language Analysis Through Smartphone Keyboard Logging} %by Timo
\title[Putting Language into Context]{Putting Language into Context Using Smartphone-Based Keyboard Logging} %by Timo
%\shorttitle{Mobile Language in Context} %"Mobile Language in Context" oder "Contextual Mobile Language"

%% Alternative A : Studying language in the wild collected with smartphones: A case study
%% Alternative A : Studying a privacy-protective data set of language collected with smartphones in the wild

\author{Florian Bemmann}
\orcid{0000-0002-5759-4976}
\affiliation{%
  \institution{LMU Munich}
  \streetaddress{Frauenlobstr. 7a}
  \city{Munich}
  \postcode{80337}
  \country{Germany}}
\email{florian.bemmann@ifi.lmu.de}

\author{Timo Koch}
\affiliation{%
  \institution{Institute of Behavioral Science \& Technology, University of St. Gallen}
  \city{St. Gallen}
  \country{Switzerland}
}
\orcid{0000-0001-6728-2063}

\author{Maximilian Bergmann}
\affiliation{%
 \institution{Institute of Behavioral Science \& Technology, University of St. Gallen}
  \city{St. Gallen}
  \country{Switzerland}
}
\orcid{0000-0002-1272-0916}

\author{Clemens Stachl}
\affiliation{%
 \institution{Institute of Behavioral Science \& Technology, University of St. Gallen}
  \city{St. Gallen}
  \country{Switzerland}
}
\orcid{0000-0002-4498-3067}

\author{Daniel Buschek}
\affiliation{%
  \institution{Department of Computer Science, University of Bayreuth}
%  \streetaddress{30 Shuangqing Rd}
  \city{Bayreuth}
  %\state{Beijing Shi}
  \country{Germany}}
   \orcid{0000-0002-0013-715X}

\author{Ramona Schoedel}
\affiliation{%
 \institution{Charlotte-Fresenius-Hochschule, University of Psychology}
 \city{Munich}
 \country{Germany}}
 \orcid{0000-0001-7275-0626}

\author{Sven Mayer}
\orcid{0000-0001-5462-8782}
\affiliation{%
  \institution{LMU Munich}
  \city{Munich}
  \postcode{80337}
  \country{Germany}}
\email{info@sven-mayer.com}

\renewcommand{\shortauthors}{Bemmann et al.}

\begin{abstract}
While the study of language as typed on smartphones offers valuable insights, existing data collection methods often fall short in providing contextual information and ensuring user privacy. We present a privacy-respectful approach - \textit{context-enriched keyboard logging} - that allows for the extraction of contextual information on the user's input motive, which is meaningful for linguistics, psychology, and behavioral sciences. In particular, with our approach, we enable distinguishing language contents by their channel (i.e., comments, messaging, search inputs). Filtering by channel allows for better pre-selection of data, which is in the interest of researchers and improves users' privacy. We demonstrate our approach on a large-scale six-month user study (N=624) of language use in smartphone interactions in the wild. Finally, we highlight the implications for research on language use in human-computer interaction and interdisciplinary contexts.
\end{abstract}




%   \includegraphics[width=\linewidth]{sampleteaser}

\maketitle

%\textbf{Operation ToDos:}
%\todo{Timo: Länge: ich finde das Paper aktuell noch recht lang, ggf. können wir an der ein oder anderen Stelle noch kürzen (insbes. am Anfang der Sections wo wir beschreiben, was wir in der Section steht - kann man mMn gerne kürzen)}
%\todo{Timo: ich schlage vor ein OSF repo mit open code anzulegen}

%\textbf{Story + Contentual ToDos:}
%\todo{Fokus auf Wiederholbarkeit - Leute sollen ihre eigene Kategorisierung erstellen können}
%\todo{holsitischer Ansatz: Logging auch mit einschließen + in android Komponente}

\section{Introduction}
Language is one of the most effective ways to gain insights into peoples minds \cite{pennebaker2017mind,chomsky1992explaining}. Designing personalized interfaces with great user experience in technical systems requires a profound understanding of user`s inner feelings and thoughts in relation to their interaction behaviors in specific contexts.
% Vorschlag Ramona für folgenden Absatz: 
Digital footprints are left behind in people's everyday language, such as Facebook posts~\cite{schwartz2013personality} or WhatsApp instant messages~\cite{verheijen2016whatsapp, koch2021whatsapp} have been shown to provide useful information on psychologically relevant traits such as personality traits or depression~\cite{schwartz2013personality, Eichstaedt2018depression}. 
Hence, understanding the mental models, attitudes, psychological states and dispositions of users is a core goal of research in human computer interaction and in neighboring fields (e.g., psychology, behavioral science). User modeling is often used in HCI to quantify individual differences for system adaption in a technical systems.
%In the latter case, language analysis is especially important, because it contributes to developing personalized mental health support~\cite{Eichstaedt2018depression}. %However, the aforementioned common approach of analyzing language data from social media platforms has one major drawback: This data usually represents only a limited snapshot of posts, which may not be representative of ``natural'' language use (i.e., private communication) and may also depend on behavioral norms of specific apps or platforms.

%\todo{P1.2. What is the specific problem?}
%\todo{P2. The second paragraph should be about what have others been doing}
%\todo{P2.3. Why is the problem important? Why was this work carried out?}
%Research thereby mostly focuses on specific aspects, e.g., interpersonal communication in one-to-one relations~\cite{koch2021whatsapp}, or how people express themselves in a semi-public context (i.e., social media)~\cite{Eichstaedt2018depression}. Furthermore, adaptive applications such as context-aware keyboards aim to adapt to the user's current circumstances, e.g., to suggest words that are appropriate to the typed type of message.
%\todo{- Related Work section - some people have done this , others that}
%\todo{However, often it is challenging to acquire the data that is necessary to allow for effective user modeling. People do not want to answer surveys, lots of details necessary to understand users, behavior needs to be understood in the context of specific apps requiring detailed information from users private phones}
Regarding language in context is thereby very important, as the meaning of textual contents highly varies with the situation in that it is expressed.
To regard text inputs into context when logging smartphone typing behavior, usually, the target app is used as a proxy, for example, by filtering for a defined set of communication apps (i.e., WhatsApp, Signal, Telegram), or using an app categorization mapping (cf. \cite{Schoedel_Oldemeier_Bonauer_Sust_2022,stachl2020predicting}). That approach has two major drawbacks: 
1) Defining a list of apps that contain the desired behavior (i.e., messaging or posting on social media) is difficult. A wide variety of apps exist %(our study showed text entries into 3325 distinct apps),
and their relevance changes rapidly.
%For example, the social media app Tik Tok\footnote{TikTok Android App \url{https://play.google.com/store/apps/details?id=com.zhiliaoapp.musically}} might not have been on a researcher's radar in a study a few years ago, while it has become one of the most relevant social media apps nowadays~\cite{feldkamp2021rise}. 
Thus, categorization approaches require frequent adaptation. 
2) An app may be used for various purposes and thus, cannot be assigned to one category. For example, the very popular app Instagram is used frequently to post public content and direct messaging. However, it is regarded as a social media app only in most app categorizations. Also, search fields that exist in nearly every app contaminate the resulting text data.

% A BLOCK ON PRIVACY. COULD MAYBE BE USED SOMEWHERE
%In this context, recent advances in mobile sensing methods offer a complementary approach as they allow researchers to collect information about participants' behavior unobtrusively in the background over longer periods of time~\cite{stachl2020predicting,harari2020sensing}. However, this comes with potential threats to participants' privacy. Especially keyboard logging, which collects all entered text in clear~\cite{aware}, can capture sensitive content~\cite{rubel2019key}. As a response, more recent language logging concepts process and abstract text data directly on the participant's device, in order to protect user privacy~\cite{bemmann_languagelogger,rodrigues2021wildkey}. This enables studies that analyze continuous everyday language data while preserving the participants' privacy: In particular, the text content is processed on the device to derive higher-level features like word categories or frequency counts. Only such abstracted features are logged, not the actual text content. In contrast to the LanguageLogger project of \cite{bemmann_languagelogger}, we integrated logging into a mobile sensing app instead of just replacing the keyboard. This approach, which we call \textit{context-enriched keyboard logging} is less obtrusive as it is integrated into another app. In addition, it yields novel kinds of context data, such as metadata of input fields and typing events on character-change granularity. 

%\todo{P3.5. What is new about your work?}
We propose to filter for  \textit{WHAT kind of content} users type instead of \textit{WHERE the content was typed}. %We enable this with our approach of logging and categorizing the input prompt texts into \textit{input motives}. This approach can be useful for various research paths from psychology, sociology, and linguistics.
%\todo{P3.4. What have you done?}
In this paper, we propose a holistic logging and preprocessing approach for smartphone typing data. We show how input prompt UI metadata can be used to contextualize language data and allows us to distinguish language contents by their input motive (i.e., posts, comments, messaging, search inputs). % instead of by the app. 
To enable researchers to leverage input prompt text for language contextualization in the wild, we developed (a) a category mapping that assigns input prompt texts to language motives, and (b) an Android module that can retrieve mobile text inputs in-the-wild and on-device assigns them their language motive.

%\todo{P4.6. What did you find out? What are the concrete results?}
We base our categorization on the dataset from a six-month representative study (N=624) that is the first of its kind using this approach. 
%\todo{P4.7. What are the implications? What does this mean for the bigger picture?}
With this context-enriched keyboard logging and analysis method, we show the untapped potential for research in sensing language data. We publish our input prompt motive mapping, alongside an Android library that allows other researchers to reuse our system in the wild. We critically discuss the threats and dangers that are posed by such technology. While our logging approach already improves user privacy, we point out space for further improvements and motivate future work in that area.

% Wie es weitergeht
% paper organization
%This paper is organized as follows: In \autoref{sec:rw} we show how mobile language data has been collected and contextualized so far, and explain how interdisciplinary research could benefit from regarding input prompt texts. We present our idea and approach of input prompt text categorization in \autoref{sec:approachoverview}. In \autoref{sec:study} we describe the field study that we conducted to collect data, which we then used in \autoref{sec:categorymapping} to create the input prompt text categorization process into input motives and give an overview of the resulting data in \autoref{sec:descrres}. In \autoref{sec:discussion} we finally discuss opportunities for interdisciplinary research and privacy that are enabled by our approach, and also talk about its limitations and appropriate future work.

\section{Related Work}\label{sec:rw}
%We contrast related work on collecting language data from digital footprints, for example, from social media and instant messaging to new data logging collection methods using smartphones. We highlight the limitations of existing methods and lay out how those can be overcome by a privacy-respectful, on-device abstraction approach, as used in our study. Furthermore, we give an overview of the past approaches towards contextualizing text input data and contrast it to our presented research method, explaining why and in which cases our approach provides advantages.

%\subsection{Collecting Language Data from Digital Footprints}

\subsection{Analyzing Language Data from Digital Footprints}\label{sec:smartphonelanguagedata}

%\subsection{Analyzing Language for Research}
Analyzing language use offers a window into people's thoughts, feelings, and motivations. Consequently, the analysis of language is a central topic in many research fields. Initially, rather unstructured language analysis methods were replaced step by step by automated text analysis tools in the last century~\cite{boyd2021}. The general idea behind all methodological approaches is to quantify raw language data.

% 1) Es wurden bisher text Daten von Social media analysiert
% SM
To investigate human language use in the wild, researchers across many disciplines have started to collect digital footprints. Many of these prior studies relied on social media text data since it is easily accessible: Researchers collected existing content by crawling web pages~\cite{Tausczik2012, Yarkoni2010}, through an Application Programming Interface (API), (such as Twitter and previously for Facebook~\cite{Golbeck2011, schwartz2013personality}), or manual content collection~\cite{Barak2007}.
% IM
An alternative source of digital footprints of text data is instant messaging services such as WhatsApp or Telegram. Private instant messaging is growing in popularity and increasingly moving into the focus of tech companies ~\cite{goode2019}.
Users produce immense amounts of text data using these services.
% In contrast to social media platforms, messages are sent more frequently than posts, consequently resulting in fewer gaps in the information stream.
Accordingly, previous studies indicate that users engage in more self-disclosure in private instant messages leading to more informative content ~\cite{koch2021whatsapp, bazarova2014}. 

% 2) diese kann man mit open versus closed vobalury approaches betrachten;
In this context, two approaches can be distinguished: Open vocabulary approaches derive features from the data itself (e.g., calculating frequencies of words used in a given text in relation to all used words), while closed vocabulary approaches use apriori defined dictionaries~\cite{eichstaedt2020}. For example, in psychological research, one well-established dictionary is the Linguistic Inquiry and Word Count (LIWC)~\cite{liwc2015}.
% The LIWC software computes the share of the words from its theory-derived linguistic categories, for example, positive emotion words or pronouns, in a given text. LIWC and other text analysis methods have been applied to a broad range of text data types, such as essays, novels, or news articles. For many research questions, it is thereby important to regard language from specific user situations only (for example \citet{koch2022age} requiring language from instant messages only), which is hardly possible with current language logging methods.

% 3) Nachteile
%The collected language data was further enriched with questionnaire data, by asking the participants to submit that data manually to researchers~\cite{Yarkoni2010}, or by requesting it via an app incorporated into social media platforms~\cite{kosinski2013private}.
However, text data collected from social media comes with downsides: People might less self-disclose on social media ~\cite{koch2021whatsapp, bazarova2014}. This could lead to systematic information gaps in the extracted digital footprints and, thus, to various biases (e.g. \cite{jurgens2020two}). These users, in turn, may have certain characteristics that make them more likely to disclose information about themselves on social media. For example, more open individuals are more likely to be active on social media~\cite{Correa2010}. Another plausible assumption would be that people only post on social media when they are in a certain mood, having language reflecting momentary and situational states. However, little empirical evidence for this assumption exists at this point.

Collecting instant messaging data is a challenging task for research purposes, however, because there is no application programming interface (API). Workarounds encompass manually collecting exported chat protocols separately from each user. However, this is time-consuming and may include text data from respective, possibly non-consenting, chat partners. Alternative approaches currently practiced in research are, therefore, to advise participants to send messages through the researchers' system, which then forwards them to the chat partner~\cite{song2014collecting} or adding a researcher's bot to group conversations~\cite{fullwood2013emoticon}. These practices have in common that participants may be biased in their choice of donated chats and constant awareness of being observed. Consequently, donated messages show only a limited snapshot of participants' language use, which is also again limited to only one specific instant messaging application. As a consequence, there is currently only a handful of instant messaging corpora available~\cite{verheijen2016whatsapp, ueberwasser2017s}.

\subsection{Methods for Collecting Language Data Using Smartphones}
% genereller Abschnitt über Smartphone Sensing
Another alternative to the two approaches mentioned above is using smartphones as data collectors. Smartphones are widespread because they are relatively inexpensive and have become our daily personal companions~\cite{Harari2016, Miller2012}. Sensors and APIs available in these portable devices unobtrusively collected various types of usage data in the background~\cite{Harari2016}. Combined with self-report measures collected via surveys or experience sampling~\cite{niels2017esm, Buschek2018}, they provide very attractive data sets for researchers, which has led to mobile sensing taking off in various application fields in recent years~\cite{stachl2020predicting,harari2020sensing}. Accordingly, smartphones have even been declared as a great resource for knowledge about human behavior in recent times~\cite{Montag2016}. Mobile sensing has so far had a strong methodological focus on extracting behavioral (e.g., app usage, calling) and situational (e.g., GPS, ambient light) information~\cite{Harari_Gosling_2023}. The application for collecting language data is so far a largely untapped potential of mobile sensing. In contrast to collecting content from specific applications, smartphones provide comprehensive behavioral data across all applications. In general, the literature distinguishes three approaches to collecting text data via smartphones, which will be briefly described in the following sections.

%2.3.1 - 2.3.3 nicht aufgliedern sondern in einen Fließtext und kürzen: nur knapp
% - was waren due Ansätze bisher
%    custom keyboard
Smartphones allow the user to install a custom keyboard\footnote{\url{https://developer.apple.com/documentation/uikit/keyboards_and_input/creating_a_custom_keyboard}}\textsuperscript{, }\footnote{\url{https://developer.android.com/guide/topics/text/creating-input-method}}, and thereby allow the developer to intercept the typed keys and log the users' keystrokes~\cite{rodrigues2021wildkey,lu2016learning,bemmann_languagelogger}. This rather low-level data is rich and detailed: Researchers can comprehend the writing process in every detail, and the keyboard can also listen to other kinds of events related to text input like the usage of word suggestions and auto-completion events~\cite{bemmann_languagelogger}.
%    APIs
Device APIs (Application Programming Interface)~\cite{ofoeda2019application} offer application developers easily accessible functionalities or data of the device operating system. %Frequently used APIs, for example, provide the developer with location data that is intelligently fused from multiple data sources (GPS, cellular data, ...) by the operating system\footnote{Android Fused Location Provide API \url{https://developers.google.com/location-context/fused-location-provider}}\textsuperscript{, }\footnote{iOS Core Location \url{https://developer.apple.com/documentation/corelocation}}, or provide aggregated data on app usage statistics\footnote{Android Usage Stats Manager \url{https://developer.android.com/reference/android/app/usage/UsageStatsManager}}.
% Specific stuff on the Accessibility service API that we used
On Android smartphones the Accessibility Service API\footnote{\url{https://developer.android.com/reference/android/accessibilityservice/AccessibilityService}} allows to access typed language~\cite{lee2021systematic,kalysch2018android}. While not possible on iOS, on Android, a researcher's app installed on the participants' smartphones can subscribe to text input events, and thereby reconstruct the typed content.
%    Screenshots
An alternative approach to recording comprehensive smartphone behavior that can also be used to access typed language is to take screenshots.
\cite{brinberg2021idiosyncrasies} deploy an app that takes screenshots every 5 seconds and presents a methodology using text and image recognition to extract behavioral data.
% advantages:
The screenshot approach allows for the extraction of meta data, but requires additional processing through recognition models. For example, the chat partner's name can be extracted from the chat app's header in a messaging app.

% - und was sind die Nachteile 
%    custom keyboard
%However, the downsides of this method arise from the keyboard replacement. Smartphone keyboards in the wild vary a lot, i.e., many manufacturers ship their devices with a custom keyboard. 
The downsides of a replaced keyboard are a different look and feel than the one the users are used to, what might, deliberately or unconsciously, influence their behavior. %For example, users might type slower on a keyboard they are not used to or behave differently due to the awareness of being observed. Furthermore, due to practical limitations in the development, a researcher-developed keyboard does usually not meet the performance of intelligent features of the original ones, such as word suggestions and auto-completion algorithms. \cite{bemmann_languagelogger} assessed the influence of their keyboard on participant behavior. 
Users reported a high awareness of the data logging but no induced content change. Instead, they report an influence of the UI differences on the interaction behavior \cite{bemmann_languagelogger}. %, especially due to the different auto-completion algorithms, visual UI differences, and a different way to access special characters.
%    APIs
%    Screenshots
Screenshot approaches are in practice often accompanied by a certain inaccuracy in the data. Machine Learning and text extraction approaches have to be used to extract the desired contents, which do not ensure perfect accuracy~\cite{chiatti2017textextraction}. Smartphone screenshots are heterogeneous, and it contextualizing text contents from screenshots is not trivial (e.g. distinguishing between typed and received messages in a chat history). % To obtain a nearly exhaustive record of user behavior would require high-frequency screenshots, leading to a vast amount of data, impacting smartphone performance, and battery life, and introducing challenges to the researchers regarding data management. 
Furthermore, even a fixed-frequency screenshot approach cannot reach the level of detail and comprehensiveness that an event-based logging approach does. %Manual labeling becomes unfeasible at all.
% privacy
All methods have in common that they expose participants to serious privacy risks. Raw user data is collected on a central server of the researcher and may contain privacy-invasive content (e.g., photos, illegal activities). Especially screenshots furthermore contain way more information than is actually needed and thereby contradict the principle of information minimization.

\subsection{Privacy-friendly Mobile Language Logging in Context - Leveraging Typing Meta Data}

\begin{figure}[t]
    \centering
    \includegraphics[width=\linewidth]{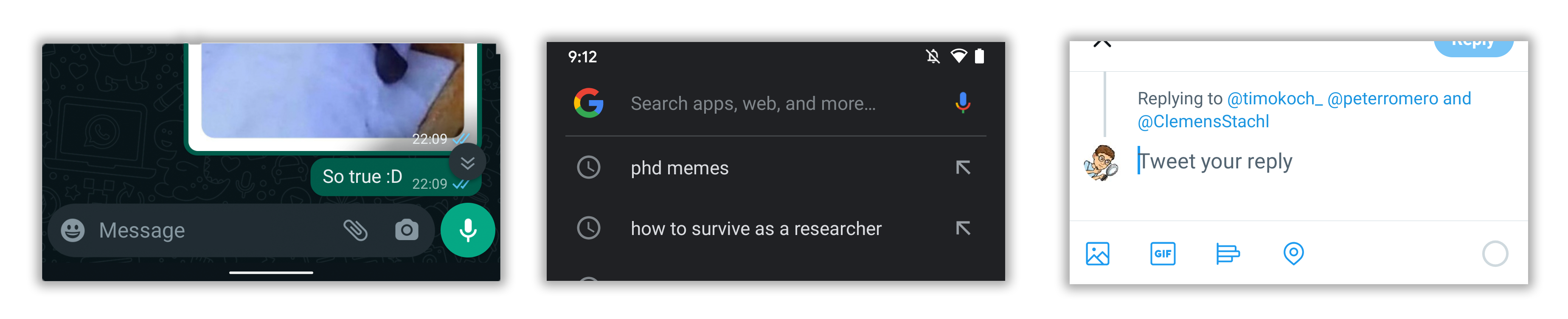}
    \caption{Screenshots of three text fields of the three Android apps: Google WhatsApp (left), Google Search (middle), and Twitter (right). All three text fields have input prompt texts, that give the user a hint about what the text field is intended to be used for.}
	\label{fig:fieldhinttexts}
\end{figure}
%Studying language use in the wild still is a difficult endeavor. To collect mobile language data, researchers replace the participants' smartphone keyboard with an intercepted one~\cite{bemmann_languagelogger,rodrigues2021wildkey,lu2016learning}, or use device APIs~\cite{lee2021systematic,kalysch2018android}. Studies on specific contexts such as language in social media mostly work with crawled (e.g. \cite{saee2017crawling,purohit2015crawling}) or user-exported \cite{boeschoten2022framework} text bodies. However, none of their approaches is satisfying yet - replaced keyboards make it hard to contextualize the content (i.e. where and in which context was it typed), and its visual differences affect user behavior and raise constant awareness on the logging. 

Besides the final text content, which is the data of interest for most applications that leverage mobile language data, a typing user also generates metadata. Depending on how the data is logged, this data may not be accessible and discarded during preprocessing. %Approaches that log the final text instead of raw input events, such as the screenshot approach of \cite{brinberg2021idiosyncrasies} or intercepted communication transmission such as \cite{song2014collecting} only access the final text content.
Implementations that access the input events by replacing the keyboard with a custom implementation instead allow access to typing metadata (cf. \cite{bemmann_languagelogger,wampfler2022}).
%Existing approaches offer a partially overlapping / partially complementary set of metadata such as touch dynamics, key event timings, or the hint text of the targeting text entry field.
However, only a few works have been conducted that make use of this metadata, and the derivation of higher-level contextual features that are relevant to interdisciplinary research is rare. Nowadays, when analyzing language use, researchers mostly select and contextualize their data by filtering by the used app, which is only a proxy for the type of content and the user's intention, but in practice not accurate. For example, social media apps can be used for both crafting public content and sending private messages.

% Welche Lücke wir füllen
%\subsubsection{Leveraging Input Prompt Texts to Contextualize Language Data}

We fill this gap and motivate further research by proposing a context-enriched keyboard logging approach.
Choosing a language logging approach that exchanges the keyboard with a custom implementation %such as \cite{bemmann_languagelogger,wampfler2022}
or leverages device APIs, meta information on where a user did type can be accessed. We specifically regard the input target app and the input prompt text, i.e., the text that is visible as a background or placeholder inside a text field as long as the user didn't type any content. We show some examples of these in \autoref{fig:fieldhinttexts}. %While the former is also available with other logging approaches through detours and thus has proliferated, input prompt texts are used and studied rarely. Furthermore, language data can be enriched with other datasources provided by the smartphone, such as passive sensor data, smartphone usage behavior, or APIs revealing information about the user's physical activity \cite{fohr2019context,wei2021sensor}. An overview is given in the review of \cite{cornet2018systematic}.
The usage of input prompt texts has not yet been studied as a data source for psychological research. Studied application scenarios in HCI encompass UI automation and privacy-enhancing tools, e.g., by \cite{pereira2020learning} to access a specific text field for automated UI testing or by \cite{wanwarang2020testing} to help a system understand the meaning and input concept of a UI element, respectively the latter by \cite{andow2017uiref} and \cite{huang2015supor} who use input prompt text to estimate whether a text field is intended to contain potentially sensitive contents. Input prompt texts can be either static or dynamic, i.e., being hard-coded and never changing, or vary depending on the user's context \cite{he2020textexerciser}.

We demonstrate an approach that infers a text input purpose and the user's intention, i.e., \textit{input motive}, from metadata of the respective text input field. We highlight which new perspectives on texting behavior it opens up, for which patterns of research it is used, and how it makes the gained data clearer and more focused.
% privacy
In doing so, we attach great importance to a privacy-friendly implementation of such techniques. Written text and the resulting logging data are highly privacy-sensitive. Our work supports a privacy-friendly, on-device-processing implementation of context-enriched keyboard logging in future user studies by proposing a set of features and abstraction methods.

\section{In-the-Wild Logging of Text Input and UI Contextual Information}\label{sec:approachoverview}
% This section describes the idea and approach and what we do. Should be easily graspable, sound novel, and be a catchy and important part

% what we do - this must sound very selling %
To tackle the aforementioned research gap, we study the potential for language use research that lies in UI metadata. We propose to use device APIs to log mobile typing behavior and create an approach that makes provided UI metadata usable to create contextual variables. We develop and provide (a) logging libraries for the Android operating systems to track language input data, and (b) a categorization approach for UI metadata that extracts a variable on the user motive of text input from the UI property input prompt text.

Our method puts an emphasis on:
\begin{itemize}
    \item \textit{Comprehensiveness.} Being able to regard text inputs across all apps and input methods.
    \item \textit{Context-sensitivity.} It should be possible to regard text inputs in the user's context.
    \item \textit{Reducing Observer Biases.} The logging method should have the least possible observer bias on the participant.
    \item \textit{Privacy.} Text input data contains privacy-invasive information, and should thus not simply be recorded as it is. 
    \item \textit{Replicability and Adaptability.} The presented approach should be applicable to many different kinds of research from interdisciplinary fields. Therefore we report our full research pipeline so that it is reproducible, and can be adapted to other research's demands.
\end{itemize}

\subsection{Input UI Metadata}
We logged the input prompt text alongside a text input through an Android API. This gives us access to all typing events at full granularity, and allows to retrieve UI metadata. We abstract each typed word into psychological categories on-device, to ensure that no raw text contents are logged. The available text input metadata encompasses the input prompt text, also a title, label, and styling/sizing properties. Label and input prompt texts tell users what they are supposed to enter and which type of input the app is expecting. Input prompt texts are visualized as placeholders, that indicate what the user is supposed to type, e.g., "Message" (WhatsApp) or "Tweet your reply" (Twitter) (see \autoref{fig:fieldhinttexts}).

\subsection{Deriving Contextual Feature: Input Motive}
In our approach, we solely rely on the input prompt texts. We create a categorization, that maps input prompt texts into one of 7 categories, i.e. \textit{input motives}. An \textit{input motive} describes the purpose and kind of content that a user is supposed to enter into a text field, such as search inputs, direct messaging, or public social media posts. We visualize this concept on the example of the app Instagram in \autoref{fig:rps}, list the input motives in \autoref{tab:hinttext-def}, and describe the categorization procedure in detail in \autoref{sec:categorymapping}.

\section{Mobile Sensing Field Study}
\label{sec:study}

In this section, we present our field study, which to the best of our knowledge, is the first large-scale deployment ($N=624$, 3 to 6 months) of smartphone-typed language logging in the wild that applies privacy-respectful on-device preprocessing. Its purpose for this paper mainly is to retrieve a body input prompt texts from real smartphone usage to create an input prompt categorization that maps input prompt texts to motives. This can be used as basis for future projects, where it allows to categorize data directly on-device. Furthermore we describe the characteristics of the obtained data, to give researchers an overview of what data to expect from such a study.

\subsection{Implementation}
% Implementation
To collect a solid data basis for our categorization in the wild, we developed an Android app that participants had to install on their smartphones. To access typed language data, it subscribes to content change events of input fields\footnote{Android Accessibility Event type VIEW\_TEXT\_CHANGED \url{https://developer.android.com/reference/android/view/accessibility/AccessibilityEvent\#TYPE_VIEW_TEXT_CHANGED}}.
Thereby it is notified on each change, i.e., each typed or removed character. For the preprocessing procedure we rely on the provided code of \citet{bemmann_languagelogger}, however our logging is implemented differently, as we did not adopt their approach of replacing the device's keyboard with a custom implementation. With the receiving event comes a reference to the target input field\footnote{EditText object \url{https://developer.android.com/reference/android/widget/EditText} } from which we retrieve additional metadata, namely the app name that the input field belongs to, and its input prompt text.

\subsection{Data Collection: Representative Smartphone Sensing Panel Study}

The study was not conducted exclusively for and tailored toward the specific needs of this paper. Data collection was conducted as part of a cooperation project between the authors' university and a research institute for large-scale psychological studies \textit{[organization names to be added in final version]}. Its purpose is to collect data to answer various research questions from psychology, sociology, and human-computer interaction.%, and thereby very suitable to demonstrate our research patterns. 
Mobile sensing, experience sampling, and survey data were collected during an individual study period of up to six months (from May until November 2020). A detailed description of data collection procedures is supplied in the pre-registration of the study protocol \textit{[details to be provided in final version]}. In the following description, we focus on the parts of the study that are relevant for this project. This research was ethically approved and carefully aligned with EU GDPR guidelines.

By means of a provider for non-probability-based online panels, a starting sample of 851 participants was recruited according to a pre-specified quota (gender, age, education, income, confession, and relationship status) representative of the German population in the age group 18-65. Participants were required to own an Android smartphone, on which they were asked to install our research app \textit{PhoneStudy}: The app had access to the users' text inputs via the Android Accessibility services\footnote{\url{https://developer.android.com/guide/topics/ui/accessibility/service}}.%ditionally tracked behavioral data (e.g., calling activities, app usage, music listening) and context (e.g., GPS, accelerometer, ambient light) in the background of the device. 
The research app ran continuously in the background of participants' smartphones during the study period of three to six months. Language data thereby were preprocessed with the Language Logger abstraction module of \cite{bemmann_languagelogger}, using the LIWC dictionary\footnote{LIWC is proprietary software. Its usage must be cleared with the authors.}~\cite{liwc2015} for word categorization and a German dictionary \cite{derewo2013v} as a whitelist for word frequency counting.
%In addition to continuous data collection, people were asked to participate in two experience sampling phases of 14 days each in study months 3 and 5. Here, participants received two to four short two-minute questionnaires pseudo-randomized throughout the day to capture, for example, their current mood, stress, and situational perception. Data collection was complemented by 30-minute monthly online surveys to assess a broad range of established psychological inventories (e.g., big five personality traits, satisfaction with life, depression). 
The full body of measures is reported in detail in the study preregistration (see \citet{schoedel2020basic}).% To inform our participants in detail on what information is logged, we designed the consent document as vividly as possible, a.o. by including examples of the language processing procedure. %Depending on the Android version participants were shown indicators when the logging of specific sensors happened, e.g. the ambient noise. However, we did not send any further prompts, to bias and annoy the users as less as possible. 
%The app tolerated the withdrawal of permissions temporarily, i.e. users interrupting the logging deliberately: They could withdraw permissions three times for a maximum of 7 days without compensation-wise consequences. After exceeding this, withdrawing permissions was only allowed for sequences of max. 24 hours for the remainder of the study.
Participants were compensated in stages, i.e., depending on how long and in which parts of the study they participated. They were excluded from data collection after several reminders if they revoked permissions for mobile sensing data collection several times for longer than seven consecutive study days and failed to complete two out of three monthly online surveys per study half. Participants could withdraw their consent at any time and ask for their collected data to be deleted. All log data was deleted from the client devices when the study finished, i.e. immediately before prompting the participant to choose their compensation.

\subsection{Sample}
\begin{comment}
\begin{figure}[t]
	\centering
\subfigure[]{
	  \includegraphics[width=0.25\linewidth]{figures/demo_age_merged.pdf}
		}
\subfigure[]{
	\includegraphics[width=0.3\columnwidth]{figures/demo_gender_cropped.pdf}
}
\subfigure[]{
	\includegraphics[width=0.3\columnwidth]{figures/demo_education_cropped.pdf}
}
\subfigure[]{
	\includegraphics[width=0.4\columnwidth]{figures/demo_legend2.pdf}
}
\caption{Comparison of our sample (blue bars) to the German population demographics (red bars). Age and gender distribution fit the national demographics quite well, however regarding education our sample underrepresents the less educated.}
	\label{fig:demographie}
\end{figure}

%\todo{[\#4] More comparison of gender ratio, education level, ...  }
%\todo{[\#21] Why at least 2 weeks? }
%\todo{[\#4] Further details; "clarify number of participants vs. duration and engagement"}

\begin{figure}[t]
	\centering
	\includegraphics[width=0.5\linewidth]{figures/users_in_study.pdf}
	\caption{Progression of the number of participants who were active in our study. The two drops at 90 and 180 days show the end of the 3 month respectively 6 month phase of the study. The slight but continuous dropout during the study is common for longitudinal mobile sensing studies.} \todo{zum thema Kürzen: Das hier kann z.b. weg}
	\label{fig:users-in-study}
\end{figure}
\end{comment}

The initial sample consisted of 851 users. 
%42 users had to be excluded because they did either not grant all necessary permissions or experienced technical issues. Another 44 users had been removed from the sample due to switching their primary smartphone during the study, which was not allowed by the study protocol. 
We excluded participants who did not grant all necessary permissions or experienced technical issues (42), switched their primary smartphone during the study (44), or participated less than two weeks (141).%which was not allowed by the study protocol. 
%Thus, we have 765 users who started to submit typing data. For this analysis we regarded only users who stayed in the study for at least 2 weeks (duration of the experience sampling waves). Thus, we excluded another 141 participants who left the study earlier. This leads to the final sample of $N = 624$.
This leads to a final sample of $N = 624$ available for further analyses. 
The mean age in our sample was 42.65 years (SD = 12.60). The age distribution follows the population in Germany%\textit{[country name removed for anonymity]} 
% \textit{[reference to data source removed for anonymity]}
, having one peak around 40 years and another between 50 and 60. Gender groups were balanced with slightly more male participants (our sample: 55.34\% male, 44.49\% female, .18\% other; national statistics: 49.34\% male, 50.66\% female, divers is not captured by the statistic)%\footnote{\change{\textit{[source hidden for anonymity]}}}.
\footnotemark{}.
%The highest completed education levels that participants were holding were distorted a little bit towards the middle: 
Regarding their highest completed education levels, our participants are rather well-educated in comparison to national averages.
We report a higher amount of participants holding a school degree (80.00\% of the participants hold a school degree as their highest education (national statistic: 52.58\%\footnotemark[\value{footnote}]), while nearly none of our participants does not hold any degree (0.81\%, national sample: 21.27\%). 19.12\% hold a university degree (national statistic: 26.15\%\footnotemark[\value{footnote}]). Only native Android users were considered as participants, i.e., people who use an Android device as their primary phone. No devices were handed out for the purpose of participating in this study.

%\footnotetext{\textit{[Source hidden for anonymity]}}

\footnotetext{Source: Federal Statistical Office of Germany \url{https://www.destatis.de/EN/Home/_node.html}}

\subsection{Data Analysis and Offline Preprocessing}
The data analysis was conducted on a central server with the statistics software R\footnote{\url{https://www.r-project.org/}}. No raw data was downloaded to local computers. The data analysis was divided into two steps: In the first step, we imported the data, did general preprocessing (e.g., parse timestamps, select the final sample), and grouped keyboard events by their text input. Word category occurrences were encoded in the many-hot format. The resulting data was in a table-like format, each row representing a text input. If not described otherwise, we analyzed events of type \textit{added} and \textit{changed} and discarded \textit{removed} events. The data were grouped per text input. In the second step, we categorized apps and input prompt texts. % as described in \autoref{sec:hinttextcatapproach}.

\section{Input Prompt Text Categorization: Distinguishing Language Contents by Their Input Motive}\label{sec:categorymapping}

\begin{figure}[t]
	\centering
	\includegraphics[width=1\linewidth]{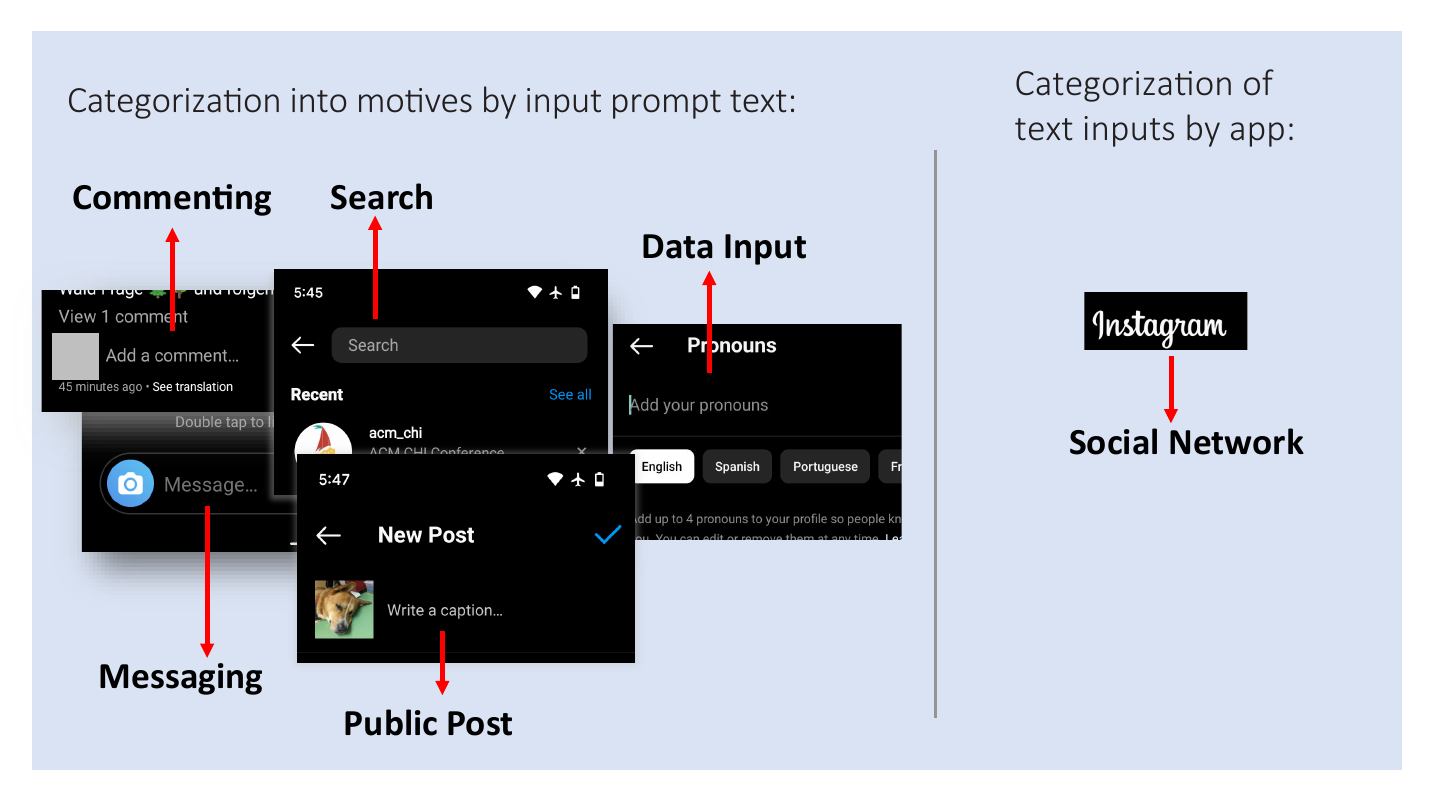}
	\caption{Instead of categorizing collected in-the-wild text input data by the originating app, we propose to regard the originating text field's input prompt text. This figure shows on the example of the app Instagram, that text inputs into Instagram are not just social media contents such as posts and comments, but can also have other motives such as messaging, search, and data input.}
	\label{fig:rps}
\end{figure}

In this section, we describe our input prompt text categorization and show the process of how we created it. We attach our input prompt text category mapping to this paper. However, we think it is important to enable researchers to create their own category mapping, to meet the specific characteristics of their dataset. We discuss the portability and validity of our yielded category mapping in \autoref{sec:valandportofcats}.

\paragraph{Prefiltering} Input prompt texts can be dynamically created by the according smartphone app, and may thus contain private content. For example, some apps extend input prompts such as "Reply to message" with the corresponding user name, i.e. "Reply to John Smith". As it is our premise not to have names of conversation partners in our dataset, we filtered such input prompt texts.
% remove texts through inter/intra user counts
We did this through the assumption, that an input prompt text that contains private information, is unlikely to occur for multiple distinct participants. Thus, we removed all input prompt texts that only occurred with one participant. We visualize this on the example of Instagram in \autoref{fig:rps}.
% TODO move to results:
%Similar to app usage logging, input prompt texts yield a long-tail distributed dataset. A few hint texts occur very often, while the vast majority occurs only very rarely. Of 22995 distinct input prompt texts in our dataset, the ten most frequent constitute 45\% of all input prompt texts.
% ca. 496000 machen die top 10 in summe, 1112317 non-NA messages gibts
To aggregate the input prompt texts into input motives for further analyses, we applied a semi-automatic, two-step categorization concept:
%In total, we categorized 4108 distinct input prompt texts (3671 of them automatically), resulting in an input motive coverage of 88.4\% of all logged input prompt texts. %We discuss limitations of our approach in \autoref{sec:lim-fht}.

\paragraph{Identification of Major Motives} We identified five major motives that users follow when composing texts on their smartphones. Those constitute our input motives: \textit{Messaging}, \textit{Posting}, \textit{Commenting}, \textit{Search}, and \textit{Data Input}. % \todo{Describe how we came up with them}
Furthermore we have the two categories \textit{Other} and \textit{Ambiguous}: input prompt texts that cannot be assigned to one of these were labeled as \textit{Other}. \textit{Ambiguous} was used if the meaning of an input prompt text was unclear at all.
We describe the input motives in \autoref{tab:hinttext-def}. Our categorization regards all smartphone text entry interactions of all participants, regardless of app category or any other pre-selection. 

\begin{table*}[t]
    \caption{\change{We categorize text inputs on smartphones into \textit{input motives}, using a text field's hint text.} }
    \label{tab:hinttext-def}
    \centering
    \begin{tabularx}{\linewidth}{lXXp{2cm}}
    \toprule
    \textbf{Motive} & \textbf{Description} & \textbf{Example} & \textbf{keywords for automatic categorization} \\
    \midrule
    \textit{Messaging} & A private message targeted to a defined person or group of people & "Type a message" (WhatsApp), "Enter your message here" (Facebook Messenger) & nachricht, message \\
    \textit{Posting} & (Semi-)public posts in social media applications. They are visible to either anyone (public posts) or limited to a group of people (e.g., friends of that user) & "Write a caption" (Instagram), "What are you doing?" (Facebook) & \\
    \textit{Commenting} & Content that is attached to an existing post, usually with the same visibility as the post & "Comment ..." (Facebook), "Tweet your reply" (Twitter) & komment, comment \\
    \textit{Search} & Content that constitutes a search query. E.g. inputs into search fields &  "Search apps, web, and more..." (Google Quicksearch), "Search photos..." (Gallery app) & such, search \\
    \textit{Data Input} & Inputs that ask the user for some information, usually form fields & "email address" (on a login screen), "Stop, address, ..." (in a public transport service app), "Spanish translation" (in a language learning app) & \\
    \midrule
    \textit{Other} & The input prompt text cannot be assigned exactly one motive, or the purpose is clear but does not belong to one of the five main motives & e.g. experience sampling and questionnaire items, "write a note..." & \\
    \textit{Ambiguous} & The input prompt text's meaning and purpose is not understandable at all & "0", "???" &     \\
    \bottomrule
    \end{tabularx}
    %  \caption{First Week}
\end{table*}

\paragraph{Categorization Step 1: Automatic keyword stem matching: }To categorize input prompt text into input motives, we applied a semi-automatic approach. In the first stage, we were looking for keywords that could be used to categorize input prompts automatically. We decided to classify all texts that contain German and respective English word snippets "such" / "search" as \textit{Search}, "komment" / "comment" as \textit{Commenting}, and those containing "nachricht" / "message" as \textit{Messaging}. With this step we could categorize 3,671 distinct input prompt texts. %We discuss limitations of our approach in \autoref{sec:lim-fht}.

\paragraph{Categorization Step 2: Manual Coding:} For further categories, we did not find a reliable matching scheme and thus, proceeded to categorize all remaining input prompt texts manually. We limited our manual categorization to only those input prompt texts which occur in more than 0.01\% of all logged texts (465 input prompt texts). The categorization was done independently by three researchers (R1 to R3), in three iterations.

% manual coding phase 1
R1 in the previous step created the category definitions and the first coding iteration was done by two other researchers independently. We thereby intend to keep the coding free of biases that arise from the motive identification phase. R2 and R3 both coded 50 randomly selected input prompt texts independently. They afterward compared their coding and discussed the differences. Their notes and suggestions were given to R1, who then improved the motive definitions. Changes included refinement of the \textit{Data Input} category, and splitting up and refining \textit{Other} and \textit{Ambiguous} (in R1's initial definition, they were joined).
% phase 2
In the second coding iteration, R2 and R3 used the refined motive definition to code the remaining input prompt texts. They agreed in 89.7\% cases, of 438 input prompts that were manually coded, R2 and R3 agreed on 393 cases, and for 45 cases their coding differed. To check the inter-coder reliability we evaluated Cohen's Kappa \cite{o2020intercoder} resulting in a nearly perfect agreement, according to the classification of \cite{landis1977measurement} ($K = 0.83$, 95\% CI: $[0.78;0.88]$).
% phase 3
The remaining discrepancies were resolved by R1 by coding the input prompts independent of the decisions of R2 and R3. Afterward the coding of R2 and R3 were additionally taken into account to overthink the coding of R1. The different coding decisions were discussed where necessary, and R1 made the final decision. This increased our final coverage to 88.4\% of all text inputs, with 4,108 distinct input prompt texts being covered.

\begin{table}
    \caption{Number of assigned motive categories, alongside disagreements and interrater agreement to each motive category of the manual coding process.}
    \label{tab:interrateragreement}
    \centering
    \begin{tabularx}{\linewidth}{X*{8}{r}}
    \toprule
    & \multicolumn{2}{c}{\textbf{Assigned Input Prompt Texts}} & & & \\ \cmidrule(l){2-3}
        \textbf{Input Motive} & \textbf{Overall} & \textbf{Manually Coded} & \textbf{Disagreements} & \textbf{Cohen's Kappa} & \textbf{95\% CI} \\
        \toprule
        Messaging & 568 & 31 & 4 & 0.88 & [0.79;0.97] \\
        Posting & 20 & 20 & 4 & 0.75 & [0.6;0.9] \\
        Commenting & 157 & 4 & 1 & 0.75 & [0.41;1] \\
        Search & 2997 & 16 & 1 & 0.78 & [0.63;0.93] \\
        Data Input & 248 & 248 & 12 & 0.85 & [0.81;0.9] \\
        \midrule
        Other & 13 & 13 & 7 & 0.8 & [0.57;1] \\
        Ambiguous & 105 & 105 & 16 & 0.82 & [0.75;0.88] \\
        \bottomrule
    \end{tabularx}
\end{table}

\section{Descriptive Evaluation of Input Prompt Text Categorized Data}
\label{sec:descrres}
% former Results section
In this section, we evaluate descriptive statistics of our input prompt text categorization. We report the characteristics of our mobile typing language dataset with input motive as a new independent variable. We show the advantages that researchers gain when selecting and filtering their language data by input motive, instead of by app category.

\subsection{The Dataset}
%\label{sec:res1}

%\begin{figure}[t]
%	\centering
%\subfigure{
%	\includegraphics[width=0.45\columnwidth]{figures/msgs_peruser.pdf}
%		}
%\subfigure{
%	\includegraphics[width=0.45\columnwidth]{figures/msgs_pertextentry.pdf}
%}
%\caption{An average user in our sample typed 23.4 text inputs per day, consisting of 146 words in total. The most prominent event type was ADDED. High standard deviations show that this highly varies among the text inputs. \todo{I somehow don't like this chart}}
%	\label{fig:descr-msgswords}
%\end{figure}

%- wie viele Daten sind zu erwarten, wie ist das so verteilt? -> Tabelle 1 links
In total, our analysis encompasses 1,868,416 text inputs across 624 users.
% --- per user level -> left subfigure ---
The average smartphone user in our sample typed 23.40 text inputs per day ($SD=28.23$). These made up 146 words ($SD = 155$) per user per day.
% --- per text entry level ---
Regarding all text inputs, each text input consists of on average 9.44 words ($SD=30.7$). During a text input, the users added on average 8.13 words ($SD=25.6$), changed 1.31 words ($SD=6.35$), and removed 1.50 words ($SD=6.86$) words.

%- Was für Channels + Apps? -> Sunburst Chart (innen channels, App cats in der Mitte, außen Apps)
%Most of the text inputs were made in communication apps (36.2\%), followed by web browsers (28.4\%) and social media apps (7.4\%).

%\subsection{input prompt texts: Coverage and Distribution among Categories}
%%%% duplicate information (see above)
%%%% reports über die input prompt texts
%- Wie viele messages / msgs with hint text gab es?  
%Of the 1,868,416 text inputs, for 1,112,317 a input prompt text could be logged. Thereby 22,896 distinct input prompt texts occurred. In total, our categorization contains 3511 input prompt texts, covering 978,844 text inputs (88\% coverage). Thereof the automatic categorization applied for 3046 distinct input prompt texts and the manual for 465.

\subsection{Descriptives of the Input Prompt Categorisation}

Of all 1,868,416 text inputs, 1,112,317 (59.5\%) obtained an input prompt text. Our categorization labeled 983,281 (88.4\%) of them, covering 52.63\% of all text inputs.

The most frequent input motives were \textit{Messaging} (44.0\%), \textit{Search} (33.8\%) and \textit{Data Input} (12.2\%). The social media categories \textit{Posting} (1.0\%) and \textit{Commenting} (3.2\%) occurred less often. The occurrence of the remainders categories \textit{Ambiguous} (4.9\%) and \textit{Other} (0.9\%) is on a low-level, what shows that our category mapping covers the user's motives well in nearly 95\% of all text inputs.

%\begin{figure}[t]
%	\centering
%	\includegraphics[width=1\linewidth]{figures/sunburst.png}
%	\caption{This figure shows in which apps users typed how frequently, for each input motive: The inner circle constitutes the number of text inputs per input motive category, and the outer circle shows the top 5 apps in which text inputs happened per motive. \todo{improve figure, esp. labels}}
%	\label{fig:amount_words_per_action}
%\end{figure}

% Reseaerch Pattern One
\subsection{Input Prompt Motives vs. App Categories: Changes in Data Characteristics}\label{sec:res2}

\subsubsection{Number of Words per Text Input}
We compare the number of words per text input, regarding their motive category. We found that text inputs that fulfill a functional purpose, such as \textit{Search} ($M_{Search}=2.30 words$, $SD_{Search}=6.80$) and \textit{Data Input} ($M_{Data Input}=2.73$, $SD_{Data Input}=9.29$), are significantly shorter than texts of motives whose content is targeted towards other people, such as \textit{Messaging} ($M_{Messaging}=12.43$, $SD_{Messaging}=18.80$), \textit{Posting} ($M_{Posting}=12.84$, $SD_{Posting}=19.00$), and \textit{Commenting} ($M_{Commenting}=12.65$, $SD_{Commenting}=20.28$). \textit{Other} text inputs range in between of both clusters ($M_{Other}=5.32$, $SD_{Other}=9.42$). A Kruskal Wallis rank sum test with a consecutive Dunn's test (p-values adjusted with Bonferroni method) revealed significant differences between the three groups \textit{Search} and \textit{Data Input} vs. \textit{Messaging}, \textit{Posting} and \textit{Commenting} vs. \textit{Other} for all pairwise comparisons with $\alpha<.01$.
%\todo{Describe this better: bullet pooints / paragraphs for the 3 groups / clusters that we found}

\begin{figure}[t]
	\centering
	\includegraphics[width=.6\linewidth]{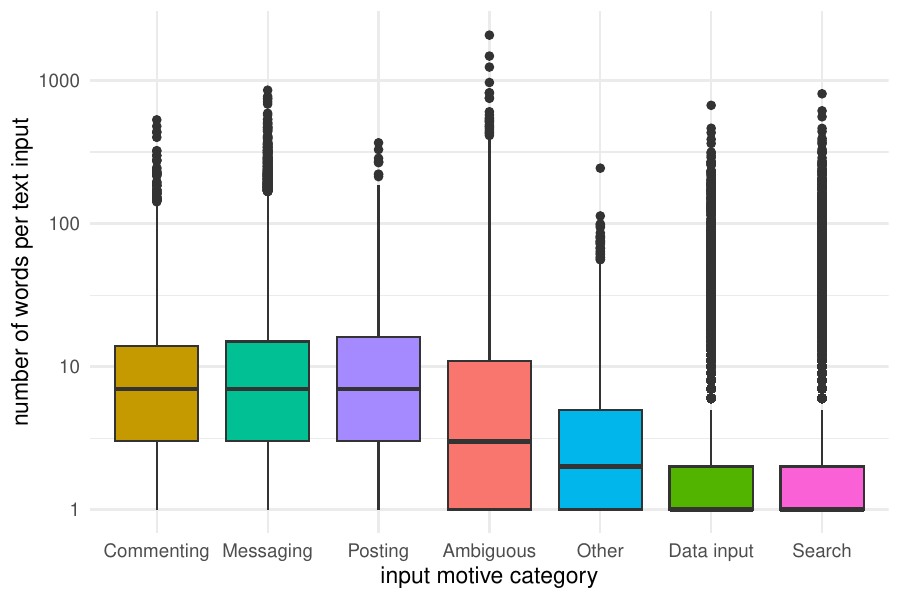}
	\caption{Words typed per user per input motive. Search inputs are rather short (1 to 3 words), and Messaging inputs are rather long with 5 to 50 words. Social network contents like posts (Content Creation) and Comments range in between.}
	\label{fig:amount_words_per_action}
\end{figure}

\subsubsection{Dictionary Matching Rates}\label{sec:res-cleninness}

%\todo{[\#1] The paper would benefit from improving the evaluations showing the coolness of pattern 1 (see idea above). At least improve the reporting of existing tests}

\definecolor{colormotives}{RGB}{242,229,203}
\definecolor{colorapps}{RGB}{200,230,227}

\begin{table}[t]
   \caption{Comparing characteristics of text inputs filtered by input motive (yellow background) and app category (green background). We compare mean and standard deviation for the two variables \textit{ matching rate} and \textit{number of words per text input}.} % Emojis dazunehmen
    \label{tab:matchrates}
    \centering
    \begin{tabularx}{\linewidth}{X*{8}{r}}%Xd{3.2}d{3.2}d{3.2}d{3.2}d{3.2}d{3.2}}
    \toprule
    & \multicolumn{2}{c}{\textsc{LIWC matching rate}}  &
    \multicolumn{2}{c}{\textsc{words per text input}} \\
    \cmidrule(r){2-3}\cmidrule(l){4-5}
    \multicolumn{1}{l}{\textbf{Input Text Selection}} & \multicolumn{1}{c}{\textbf{M}} & \multicolumn{1}{c}{\textbf{SD}} & \multicolumn{1}{c}{\textbf{M}} & \multicolumn{1}{c}{\textbf{SD}} \\
    \midrule
    \rowcolor{colorapps}App Category: Communication & 48.32\% & 31.05\% & 16.43 & 38.30  \\
    \rowcolor{colormotives}input motive: Messaging & 50.64\% & 30.33\% & 12.43 & 18.80 \\
    \midrule
    \rowcolor{colorapps}App Category: Social Media & 31.59\% & 33.52\% & 12.87 & 39.60 \\
    \rowcolor{colormotives}input motive: Posting & 38.82\% & 30.34\% & 12.84 & 19.00 \\
    \rowcolor{colormotives}input motive: Commenting & 41.78\% & 30.35\% & 12.65 & 20.28 \\
    \midrule
     \rowcolor{colorapps}App Category: System & 18.78\% & 31.67\% & 4.34 & 20.00 \\
     \rowcolor{colormotives}input motive: Search & 13.02\% & 28.60\% & 2.30 & 6.80 \\
    \bottomrule
    \end{tabularx}
\end{table}

In this section, we regard how much of the logged text contents could be matched with the applied dictionary. We analyze the different matching rates 
%we compare the matching rates 
of text input (1) selected via our motive categories derived from the input prompt texts, and (2) filtered by the app categories of \cite{Schoedel_Oldemeier_Bonauer_Sust_2022}. 
Therefore, we compare the matching rates of the extracted LIWC categories for the input motive \textit{Messaging} with the app category \textit{Messages}, for the input motives \textit{Posting} and \textit{Comment} with the app category \textit{Social Network}, and the same named input motive \textit{Search} with the app category \textit{System} (a comparison with an app category is not clearly possible in this case. We have chosen the category System, as it contains some search apps%in the used app categorization, search apps are contained in the category System
).

In general, we found matching rates for messaging content being the highest (around 50\% for both kinds of categorization), followed by social media content between 30\% and the lower 40s. Search inputs, in general, match the dictionary rather badly (10\% to 20\%). This characteristic is not surprising, as messaging consists of rather natural language, whereas hashtags and more net-speech might enhance social media language. Search queries contain rather specific words (names, locations, etc.) that are not contained in the dictionary.

Filtering content by input motives instead of app categories could strengthen these characteristics: We found higher matching rates for messaging and social network content when selecting text input via input motives than when using the equivalent app categories (please find the effect sizes reported in \autoref{tab:matchrates}). For the category \textit{Search}, the matching rate behaves oppositely, i.e., the \change{input motive \textit{Search} shows a lower matching rate than the respective app category \textit{System}}. Standard deviations when filtering by motive category instead of app category were lower in all comparisons. %\todo{kann man sagen dass das für bessere Daten spricht?}

%\idea{[\#1] Idea 1: Show that effect sizes increase when filtering by content type. E.g. Timo did some analyses that showed that effect size of correlations between affect and use of I related words increase when looking on public content instead of all inputs}

%\idea{or some clustering to show data "cleansity"}

\section{Discussion}\label{sec:discussion}

In this section, we discuss our context-enriched keyboard logging approach. We interpret which implications the results of our data analysis have on research patterns, how they can be used to foster future studies by other researchers, and which dangers and privacy issues are revealed.

\subsection{More Specific Data by Comparing Language Across Communication Channels}
\label{sec:disc-clearerdata}
The analysis of our study data showed that selecting text inputs via the input prompt text yields higher matching rates in the LIWC dictionary than when selecting via app categories for messaging and social media content. % (see \autoref{sec:res-cleninness}).
Although we did not compare modeling scores of exemplary dependent variables and psychological constructs (such as predictions of age and gender \cite{koch2022age}, affective states \cite{koch2022affect}), this implies that the yielded data is cleaner and of higher quality when selecting text inputs via the input prompt text.
The opposite behavior for the input motive \textit{Search} is not surprising and can be explained by the characteristics of search inputs: They are very short, usually just a few words (see \autoref{fig:amount_words_per_action} for comparison) and consist of specific terms (companies, names, locations, ...). They are, in general, more ambiguous and, thus, often not covered by closed-vocabulary approaches, such as the LIWC dictionary. Thus, lower matching rates for the input motive \textit{Search} in comparison to the equivalent app category are not a bad sign but instead speak for a cleaner body of actual search inputs. To study search inputs using the presented approach, we recommend designing a dictionary specialized on the underlying research question first, such as \citet{remus2010sentiws} created for sentiment, or \cite{cheng2016psychologist} with who collected keywords indicating depressive states.
%The analysis of message lengths also supports the usage of input prompt texts resp. input motives to filter text inputs. Lower standard deviations for text inputs selected by input motives in comparison to their app category equivalents denote more homogeneous data.

% and now the critical stuff:
While the results of our study indicate that selecting text inputs by input motives instead of app categories yields higher data quality, more research is needed to test this, empirically. %we did not prove this statistically. 
We motivate future work to assert the data quality: Language motives could be regarded as clusters and evaluated clustering evaluation metrics such as inter-cluster and intra-cluster distances (c.f. \citet{ranby2016comparison}), or more language-specific similarity metrics as evaluated by \citet{krakovsky2013clustering}. Also, it would be interesting to redo analyses of existing research on mobile language, with both app-categorized data and input prompt text categorized data, to see the effect on the results, for example, improvements in effect sizes.

% THE STUFF ON 4 LEVELS OF CONTEXT
%Connecting language data with contextual information from input prompt texts enables novel research opportunities.
%\cite{boyd2021} introduce four layers of context, in which words are observed in social psychology. Our methods enables to regard the social context, which is subdivided into \textit{acute situation} and \textit{environment}. It can be taken into account via the input motive, i.e., into where a word is typed.

%Applying our research patterns, it is possible to regard all four levels in psychological models: Word sequences as proposed in \ref{sec:rp3} map the \textit{discourse} context (i.e., word meaning). The \textit{person} context can be regarded by assessing traits via questionnaires or between-user modeling. The social context, subdivided into \textit{acute situation} and \textit{environment}, can be taken into account via sensing data and the input motive, i.e., into where a word is typed.

\subsection{Privacy}% - Risks and Dangers}
%\todo{[\#16] Good callout here, extend this discussion}

%\todo{kritische Betrachtung von Sprach Logging}
With great power also comes great responsibility. While the availability of contextual information and language data yields many opportunities, it also creates privacy risks. While people are mostly aware of the risks of language data and have concerns about sharing them \cite{bemmann_languagelogger}, awareness of risks about what can be inferred from passive sensing data is not at an adequate level yet \cite{harari2020process}. Thus, using methods and tools like the ones presented in this work, introduces huge responsibilities to the applying researchers: The preprocessing features are privacy-wise only as strong as their configuration. In the privacy-wise worst-case application, one could configure the preprocessing with a one-to-one mapping of all words contained in a language dictionary to the same word. The content abstraction approach would thereby be undermined. This issue, however, is independent of our input prompt text categorization approach.

%\subsection{Privacy - Opportunities}

%Besides the increase in privacy risks that the contextual enrichment of language data brings (c.f. the previous subsection),
More precise pre-filtering of text contents, that our method enables, does support user privacy.
%Filtering text contents by input prompt text allows for more accurate pre-filtering. 
In line with the recent calls for data minimization (c.f. \cite{hatamian2020engineering}), contextual information on text inputs allows for discarding unnecessary data immediately on-device. An estimation on the example of studying social media post language on Instagram shows that our approach allows removing unnecessary text inputs even before they leave the participant's smartphone. Of 38,075 text inputs into the app Instagram, only 4693 inputs reflect posted public content (posts and comments), whereas 9992 inputs were public messages and 154 data inputs. As the latter two are not relevant in regard to public social media language, but might contain very privacy-sensitive content, they should be excluded for the sake of data minimization. Our approach allows us to do so directly on the participant's smartphone.

%Besides extending the possibilities for researchers in various fields, that we depicted in the previous sections, input prompt texts can also be used for data minimization, i.e., to improve user privacy. By being able to filter for specific text inputs more finer, developers can limit the data logging more precisely to what is really needed. However the presented methods individually are not new privacy-wise and, without meaningfully connecting them, do not automatically improve users' privacy. Also, as each method does, ours poses its individual set of risks and vulnerabilities (e.g. data being temporarily stored on the client device). However, with a meaningful combination, the amount of privacy-sensitive data that has to be processed to answer a research question can be reduced. More fine-grained filtering applying multiple filters at once (e.g. app category, input prompt text, sensing-based information like the user's situation) reduces the amount of data, while a far-reaching preprocessing that applies well-fitting dictionaries reduces the sensitivity of the actually logged information.

% - context abhängige logging settings
%More fine-grained logging can also be achieved by regarding passive sensing data. Depending on the logging purpose, it can be configured to become active only in specific situations, e.g., activities or locations.

\subsection{Future Work: Derive Further Information from UI}

Our work is, to the best of our knowledge, the first that investigates the usage of the input prompt text as contextual variable about the user for research. Building on this starting point, we see more potential in input prompt texts that should be explored in future work:

% WEitere Kontextinfos aus Meta Daten: z.B. Sprache der App durch input prompt text
\paragraph{Derive context from input prompt text} In our work, we used the input prompt texts to categorize text inputs by the kind of input that developers expect. However, other kinds of categorizations are possible as well. For example, the input prompt text can be used to infer the language an app is configured to be in, which likely also is the input language.

%Auch labels auswerten
\paragraph{Take surrounding UI into account} Besides the input prompt text, surrounding UI elements could also be of interest, such as the label of a text field or the input field's size.

\paragraph{Automatic Categorization of Yet Unseen Input Prompts} Our approach only works for known input prompts, i.e., ones that are included in the category mapping exactly as they are. To enhance the procedure's coverage, approaches that can deal with yet unseen input prompt texts would make an improvement. That could be the matching of word stems and keywords, word embeddings, or deep-learning-based approaches.

\subsection{Limited Availability of Input Prompt Texts}
An input prompt text is not available with all text fields. Our implementation could access input prompt texts for 59.5\% of all typing sessions. The reasons are that input prompt texts are not always set by the app's developer, and that, for some cases, we cannot distinguish between input prompt text and pre-set text, and thus have to discard the content for privacy reasons. Furthermore, access to accessibility services might become more difficult with future Android versions. To enhance the coverage of information, other UI properties and elements, such as labels, could be taken into account. They could be processed with the same procedure as input prompt text, as they fulfill a similar purpose. 

As mobile operating systems generally are a matter of change, our approach might need adaptation for future deployments. Although our implementation adhere's to current Android policies, the availability of the data that we rely on may change in the future.

\subsection{The Long-tail Distribution of Categorization Coverage} 
The proposed categorization approach is not holistic. In our study, we cover 88.4\% of all logged input prompt texts, covering 52.63\% of all text inputs. Input prompt texts occur in a long-tailed distribution, i.e., the top 10 most frequently occurring input prompt texts (0.0004\%) occur in 496,135 input prompts (40.60\%) of our text inputs. Thus, coding the remaining uncovered 11.6\% requires a way higher coding effort than the first 88.4\%. 
Furthermore, input prompt texts change with apps evolving, or new apps coming up. While the automatic coding approach might cover most of the new input prompt texts, some others will not be covered without updating the manual coding. This issue is already known from app categorizations, which also need to be updated frequently as new apps proliferate at a high speed. However, due to the semi-automatic approach, we expect our input prompt text approach to be more stable for changes than nowadays app categorizations are. With 22,896 distinct input prompt texts being covered already a significant proportion of the input prompt texts of new apps might already be known, and another huge proportion might be covered by automatic categorization. App categorizations, in contrast, have to deal with each new app individually.

\subsection{Portability and Reusability of our Categorization}\label{sec:valandportofcats}

%\todo{subseciton of selectivity of the 53\% coverage. Auch interessant: selectivity so ener panel studie: sehr hoher anteil von texteingabe in com.nielsen.consumerpanel}

% I deliberately excluded validtiy aspects, as they are not that relevant in ACM 
The input prompt text categorization that we provide is fitted to our specific study data, thus to what extend this generalizes to new data collections is unclear.
%It only works for input prompt texts that were included in our sample's mobile typing data.
Other studies' participants might use yet unseen apps, and input prompt text within yet-known apps might change with app updates and thereby the number of unclassifiable input prompt texts might increase.
Due to our large-scale and close-to-representative sample encompassing 3,325 distinct apps with 22,895 distinct input prompt texts, we expect our categorization to be rather stable and well-usable in the next few years. However, the issue of a categorization becoming outdated over time, which is well known with app categorizations (c.f. \cite{Schoedel_Oldemeier_Bonauer_Sust_2022}), is well known for closed-vocabulary approaches, and does also apply to ours. Especially when deploying our approach in a specific context, i.e. with a specific kind of participants, country, or regarding a specific rarely used apps, our categorization might not yield a satisfying coverage.
Therefore, not just the deployability of our categorization but also the reproducibility of the categorization process itself is very important to us.
We recommend researchers pilot the classification with a small subsample of participants before applying it fully on-device. If necessary, one can then either extend the classification before the full deployment or opt for an offline categorization approach that initially keeps the raw data. %We give information on how to deploy the existing categorization in \autoref{sec:deploycats}, and show how the categorization can be extended or redone on \autoref{sec:redocats}.

\section{Conclusion}
In this paper, we motivate researchers who work with mobile language data to take contextual data into account. Distinguishing language contents by their input motive (i.e., comments, messaging, search inputs) creates a yet-unused contextual variable. We show how smartphone app UI's input prompt texts can be preprocessed on-device, and provide an input prompt text categorization. We base our work on a representative large-scale six-month user study (N=624) of language use in the wild. With the analysis of our study data, we show which data characteristics researchers have to expect from such a study and give descriptive insights into our contextual variable \textit{input motive}. We argue that context is beneficial in many ways: It allows for better prefiltering of mobile language data than the currently prominent approach, selecting texts by app, does. This enables a more fine-grained selection of input texts and improves user privacy, as unnecessary content can be discarded earlier. Finally, we highlight implications for research on language use in psychology, linguistics, and interdisciplinary contexts. Novel research paths are enabled that support future work on mobile language use in the wild. Besides the advantages and use cases, we also discuss the imposed risks on dangers that such data have on user privacy.

\begin{acks}
%\todo{ZPID for paying, Bavarian state for funding Daniel}
We would like to thank the Leibniz Institute for Psychology (ZPID) for their great cooperation and support in the Smartphone Sensing Panel Study, from which the dataset published here arises.
\end{acks}

\bibliographystyle{ACM-Reference-Format}
\bibliography{bibliography_apa7}

%\appendix

%\section{How-To Reuse this for Future Studies}

%\subsection{As-it-Is Deployment}\label{sec:deploycats}
%\todo{can be used as it is. Getting started quickly.}

%\subsection{Re-doing the Categorization}\label{sec:redocats}

%\todo{for some purposes is mithgt be necessary to redo the categorization}

% \begin{figure}[t]
% 	\centering
% 	\includegraphics[width=.8\linewidth]{figures/liwcmatch_per_action_heatmap.pdf}
% 	\caption{Amount of LIWC categories per action. Values are normalized by the amount of words per action. {\color{red}TODO im Text erwähnen}}
% 	\label{fig:liwc_match_action_heatmap}
% \end{figure}

% \begin{figure}[t]
% 	\centering
% 	\includegraphics[width=.8\linewidth]{figures/liwcmatch_per_action_heatmap_2.pdf}
% 	\caption{Amount of LIWC categories per action. Values are normalized by the amount of words per LIWC category. {\color{red}TODO im Text erwähnen}}
% 	\label{fig:liwc_match_action_heatmap}
% \end{figure}

\end{document}